\documentclass[letterpaper,9pt]{IEEEtran}

\usepackage{cite}
\usepackage{amsmath,amssymb,amsfonts}
\usepackage{algorithmic}
\usepackage{graphicx}
\usepackage{textcomp}
\usepackage{xcolor}
\usepackage{mathrsfs}

\usepackage{amsfonts}
\newtheorem{lem}{Lemma}
\newtheorem{thm}{Theorem}

\newtheorem{rem}{Remark}
\newtheorem{exmp}{Example}
\newtheorem{defn}{Definition}

\usepackage{indentfirst}

\usepackage{flushend}

\usepackage{multirow}
\usepackage{float}
\usepackage{algorithm}
\usepackage{algorithmic}
\usepackage{nomencl}

\usepackage{etoolbox}

\allowdisplaybreaks

\begin{document}

\begin{@twocolumnfalse}

    \title{
                \begin{flushleft}
            \Large \textbf{Letter}
        \end{flushleft}
    \vspace{0.5cm}
        \large \textbf{ Stabilization with Prescribed Instant \\via
        	Lyapunov Method  }}
    \author{Jiyuan Kuang, Yabin Gao, Yizhuo Sun, Jiahui Wang, Aohua Liu, Yue Zhao, Jianxing Liu*
\thanks{Jiyuan Kuang, Yabin Gao, Yizhuo Sun, Aohua Liu, and Jianxing Liu are with the department of Control Science and Engineering, Harbin Institute of Technology, Harbin 150001, China. (Email:  sdukuangjiyuan@163.com, gaoyb2012@gmail.com, syz-hit@hit.edu.cn, 21s104175@stu.hit.edu.cn, yue.zhao@hit.edu.cn, jx.liu@hit.edu.cn). }

\thanks{Jiahui Wang is with the College of Intelligent System Science and Engineering, Harbin Engineering University, Harbin 150001, China. (Email:  jiahuiwang@hebut.edu.cn). }

\thanks{    {Corresponding author: Jianxing Liu.}}
}

    \maketitle
\end{@twocolumnfalse}

\vspace{-1cm}\noindent Dear Editor,\\

This letter investigates the prescribed-instant stabilization problem for high-order integrator systems. In anothor word, the settling time under the presented controller is independent of the initial conditions and equals the prescribed time instant. The controller is designed with the concept of backstepping.
A strict proof based on the Lyapunov method is presented to clamp the settling time to the prescribed time instant from both the left and right sides. This proof serves as an example to present a general framework to verify the designed stabilization property.  
It should be emphasized that the prescribed-time stability (PSTS) \cite{2017prescribed1} can only prescribe the upper bound of the settling time and is different from this work. The detailed argumentation will be presented after a brief review of the existing important research.

Traditional asymptotic stability ensures the system states converge to equilibrium as time goes to infinity. Since the system states actually never reach an equilibrium, the separation principle  must be rigorously substantiated \cite{basin}. 
Finite-time stability (FNTS) guarantees that states convergence happens in a finite time but mostly depending on parameters and initial conditions \cite{2004Finite}. 
By using a finite time differentiator or observer, the correct information can be estimated after a finite time \cite{b12}. 
This makes it easier to get the closed-loop system stability.
 However, this finite time increases as the initial values of system states increase, and there are no uniform bounds. To solve this problem, one way is to estimate the  settling time in some frequently used finite-time stabilization algorithms, such as super-twisting algorithms \cite{b13}. Another way is developing some new algorithms that can ensure a uniform bound of the settling time.
The fixed-time stability (FXTS) guarantees the settling time to be
bounded by a constant, which however is not explicit and is determined by the controller parameters \cite{2011Fixed,2011Uniform}. So, it is complex to calculate
every parameter according to the desired bound of settling
time \cite{2016MIMOFixed}. Moreover, the settling time in FXTS is very conservative.

The prescribed-time stability (PSTS) ensures the system states converge to zero in a prescribed time $T_p$, where $T_p$ is an explicit parameter of the controller \cite{hefu}. 
Some results of PSTS even show a pre-specified settling time (in simulations at least) \cite{observer, raodong}. However, their corresponding theoretical analysis cannot explain this fact, except for some first-order systems. A detailed analysis can be seen in Remark \ref{defence1} and \ref{defence2}.

Up to now, only a few works with strict proofs forced the settling time to an arbitrarily selected time instant. For example,
the work in \cite{beihang} ensured this property by using a novel sliding mode control. The corresponding  proof was demonstrated through a detailed analysis of the infinitesimal order of each state.
The work in \cite{mytc} designed a controller based on the backstepping method. Reduction to absurdity was utilized to verify the exact settling time. 
However, these methods of proof are circumscribed and can not be generalized easily.

This letter considers $n$-order integrator systems, of which the settling time under the presented controller is exactly the prescribed time instant. 
A corresponding proof based on the Lyapunov method provides a general framework to verify the exact settling time. Moreover, this framework can also help to decrease potential conservativeness in the settling time of traditional PSTS.

\noindent\textbf{Problem Statement:}
Consider the following system	
\begin{equation*}
	\dot x= g(x,u),
\end{equation*}	
where $g:\mathbb{R}^{n}\times \mathbb{R} \rightarrow \mathbb{R}^n$, $x \in \mathbb{R}^n$ denotes the states, and $u \in \mathbb{R}$ is the control variable. Consider the control variable as $u(x,t,\eta)$, where $\eta \in \mathbb{R}^m$ denotes the parameters. We can obtain the closed-loop system in \eqref{ft1} with initial value $x(t_0)=x_0$. The initial time $t_0=0$ is default in this letter.  
\begin{equation}
	\dot x=f(x,t,\eta):= g(x, u(x,t,\eta)).\label{ft1}
\end{equation}

\begin{defn} \label{d1} \cite{mytc}
	If for any physically	possible positive number $T_p$, there exists parameters $\eta$ such that the system settling time $T(x_0)$ can be prescribed as $T(x_0)=T_p, \forall x_0 \in \mathbb{R}^n$. Then, the origin of the system \eqref{ft1} is said to be prescribed-instant stable (PSIS).
\end{defn}

Consider the following linear system:
\begin{equation}
	\left\{
	\begin{array}{lr}
		\dot x_i=x_{i+1}, i=1,...,n-1,  &  \\
		\dot x_n=u.
	\end{array}
	\right.\label{norderx}
\end{equation}
Since the PSIS has already been defined and proved in \cite{mytc}, the main contribution of this letter is presenting a Lyapunov method to verify the controller can ensure the system in \eqref{norderx} is PSIS.

	The expressions of Theorem 2 in \cite{2017prescribed1} and Theorem 1 in \cite{raodong} may mislead the readers to think that the PSTS also ensures $T(x_0)=T_p$. 
	To clear the air, it is urgent to emphasize the following fact.

\begin{rem} \label{defence1}

	The proof in \cite{2017prescribed1} is one of the main  thought  of proof of PSTS. The key step is to obtain the following inequality:
	\begin{equation} \label{song}
	\frac{\text{d}V(t)}{\text{d}t}\le -2 k \mu(t) V(t), \; k>0, \; t\in [0,T_p),
	\end{equation}
	where $V$ is a Lyapunov function of the controlled system and
	\begin{equation}
	\mu(t)=\frac{T_p^{m+n}}{(T_p-t)^{m+n}}, \; t\in [0,T_p).
	\end{equation}
If the formula in \eqref{song} is equality, there is no doubt that the settling time equals the prescribed time $T_p$.  However, for a high-order system, it is difficult to obtain equality of \eqref{song}. As a result, $V(t)$ is reset to zero in the prescribed time \emph{no longer than}  $T_p$ irrespectively of the initial value $V(t_0)\in \mathbb{R}_{\ge 0}$ (Section 2 in \cite{ORLOV2022}). We have $T(x_0)\le T_p$. 
A similar problem also exists in \cite{observer}, whose equation (27) in Theorem 1 is also an inequality.
\end{rem}

\begin{rem} \label{defence2}
Another thought of proof of PSTS in some research such as \cite{raodong} is based on some time scale transformation from $t \in [0,T_p)$ to $\tau \in [0,+\infty)$. The most familiar transformation  is
\begin{equation}
	\left\{
	\begin{array}{lr}
		\tau=-\text{ln}(\frac{T_p-t}{T_p}),  &  \\
		t=T_p(1-\text{e}^{-\tau}).
	\end{array}
	\right.
\end{equation}
 Since $\frac{\text{d}\tau}{\text{d}t}=\frac{1}{T_p-t}$, we have two functions equivalent to each other:
 \begin{equation} 
 	\begin{cases}
 		\frac{\text{d}V(t)}{\text{d}t}= -\frac{1}{T_p-t} V(t),,&t\in [0,T_p);\\
 		\frac{\text{d}V(t(\tau))}{\text{d}\tau}=\frac{\text{d}V(t)}{\text{d}t}\frac{\text{d}t}{\text{d}\tau}=-  V(t(\tau)), &\tau\in [0,+\infty).
 	\end{cases}
 \end{equation}
If one can prove the Lyapunov function $V(t(\tau))$ converges exponentially, or $V(t(\tau))\rightarrow 0$ as $\tau \rightarrow +\infty$, the conclusion is definitely obtained that $V(t)\rightarrow 0$ just as $t \rightarrow T_p$. However, Lemma 2 in \cite{raodong} only shows $	\frac{\text{d}V(t(\tau))}{\text{d}\tau} \le -  V(t(\tau))$. There stands a chance that $\frac{\text{d}V(t(\tau))}{\text{d}\tau} \le -V^{1.5}(t(\tau))-V^{0.5}(t(\tau))$. As a result, $V(t(\tau))=0, \forall \tau \ge \pi$.
Since $V(t(\tau))$ converges to zero before $\tau \rightarrow +\infty$,
 $V(t)$ converges to zero before $t=T_p$, i.e., $T(x_0)\le T_p$. 

Although some simulations of PSTS have obtained $T(x_0)= T_p$, we have clearly shown that the corresponding proofs of the PSTS are not sufficient to obtain this result. 
\end{rem}

\begin{rem}
	The work in \cite{2020freewill} defined the properties of PSTS (PSIS) as free-will weak (strong) arbitrary time stability, correspondingly. It recognized that any single inequality of derivative from the Lyapunov function could not obtain PSIS directly.
 However, the  free-will strong arbitrary time stability, which is consistent with the presented PSIS, can be established as long as
\begin{equation} 
	\frac{\text{d}V}{\text{d}t}= -\frac{k(1-\text{e}^{-V})}{(T_p-t)} , \; k>1, \; t\in [0,T_p).
\end{equation}
Although the  proof of free-will strong arbitrary time stability for high-order systems remains open, yet this proof can be completed once the following inequalities are considered,
\begin{equation*} 
	\begin{cases}\dot V_1 \le \frac{-k_1(1-\text{e}^{-V_1})}{(T_p-t)},&k_1>1, t\in [0,T_p);\\
	\dot V_2 \ge \frac{-k_2(1-\text{e}^{-V_2})}{(T_p-t)} 	,&k_2>k_1, t\in [0,T_p).
	\end{cases}
\end{equation*}
Specifically, one can obtain the PSIS by limiting the settling time from both the left and right sides of Lyapunov function derivative formula.  This is the main thought of the PSIS presented in this letter.
 
To realize the above-mentioned thought, one can find a differential function whose solution converges to zero just at the prescribed instant $T_p$. Definition 5 of \cite{mytc} presented a series of such functions named reference convergence differential functions (RCDFs).
\end{rem}

\begin{exmp} \label{exmp1}
	Let us see some typical RCDFs ($\psi$) given in \cite{mytc}:
	\begin{equation*} 
		\begin{cases}\dot v_1=-\psi_{v_1}=-\frac{\eta (v_1^2+1)\text{arctan}(v_1) }{ T_p-t },&v_1(t)=\text{tan}(T_p-t)^\eta ;\\
		\dot v_2= -\psi_{v_2}=-\frac{ \eta v_2 }{ T_p-t }	,& v_2(t)=(T_p-t)^\eta;\\
		\dot v_3= -\psi_{v_3}= -\frac{\eta (1-\text{e}^{-|v_3|}) }{ T_p-t }\text{sign}(v_3)	,  &v_3(t)=\text{ln}(1+(T_p-t)^\eta).
		\end{cases}
	\end{equation*}
It is noted that $\psi(v,t,T_p,\eta)$ can be written as $\frac{\eta \zeta(v)}{T_p-t}$, and $\zeta(v)$ has the same sign as $v$.  Moreover,  $\zeta(v)=O(v)$ (infinitesimal of the same order), and 
	\begin{equation*}
	\lim\limits_{t\to T_p} \psi (v,t,T_p,\eta)=\lim\limits_{t\to T_p}\frac{v(t)-0 }{t-T_p}\sim  (T_p-t)^{\eta -1}, \forall \eta>1.
\end{equation*}
 
In addition, Claims and Lemmas in \cite{mytc} provide other RCDFs may help to promote the proof of PSTS in existing research to obtain PSIS.
\end{exmp}

\noindent\textbf{Main Results:}

{\bf Controller Design:}
The controller is designed with backstepping method and is presented as a recursive form. The desired value of $x_1$ is $x_{1,d}=c$, where $c$ is a constant.
The recurrence relation ($i\ge 2$) is a little different from that in \cite{mytc},
\begin{equation}
	x_{i+1,d}=\dot x_{i,d}-z_{i-1}-\psi_i,
	\label{relation}\end{equation}
where $z_i=x_i-x_{i,d}$ and $x_{2,d}=-\psi_1$. It is noted that $\psi_i(z_i,t,T_p,\eta_i)$ belongs to the same RCDF. 
To prevent the singularity problem of the control signal at $t=T_p$,
$\eta_i$ in $\psi_i$ is designed to satisfy $\eta_i> n+1-i, i=1,2,...,n$. 
For the system \eqref{norderx}, 
\begin{equation}  u=\begin{cases}x_{n+1,d},& {0\le t<T_p},\\0,& {T_p\le t}.\end{cases}\label{ncontrol}\end{equation}
So, the dynamics of each state's tracking error when  $t\in [0,T_p)$ is:
\begin{equation}
	\left\{
	\begin{array}{lr}
		\dot z_1=z_2-\psi_1,   \\
		\dot z_{i}=z_{i+1}-z_{i-1}-\psi_{i}, i=2,3,...,n-1,   \\
		\dot z_n=-z_{n-1}-\psi_n.
	\end{array}
	\right. \label{zz}
\end{equation}

In the following, we will present the PSIS of the transformed system \eqref{zz}, and then obtain the PSIS of the system \eqref{norderx}.

{\bf Prescribed-instant Stability:}
\begin{lem} \label{qin}
	Suppose the function $h(x): \mathbb{R} \rightarrow \mathbb{R}$ is concave on the interval $I$ and $\lambda_1, \lambda_2,..., \lambda_n$ satisfy $\sum_{i=1}^{n}\lambda_i=1$. Then,
	\begin{equation}
		\sum_{i=1}^{n} \lambda_i h(x_i)  \le h(\sum_{i=1}^{n}\lambda_i x_i),\; \forall x_i \in I.
	\end{equation}
	Especially, if $\lambda_1=\lambda_2=...=\lambda_n=\frac{1}{n}$, we have
	\begin{equation}
		\frac{\sum_{i=1}^{n}h(x_i)}{n} \le h(\frac{\sum_{i=1}^{n}x_i}{n}), \; \forall x_i \in I. \label{use}
	\end{equation}
\end{lem}  

This lemma is the so-called Jenson inequality. A geometric proof is given in the following.

\begin{figure} [htbp]
	\centering
	\includegraphics[width=0.4\textwidth]{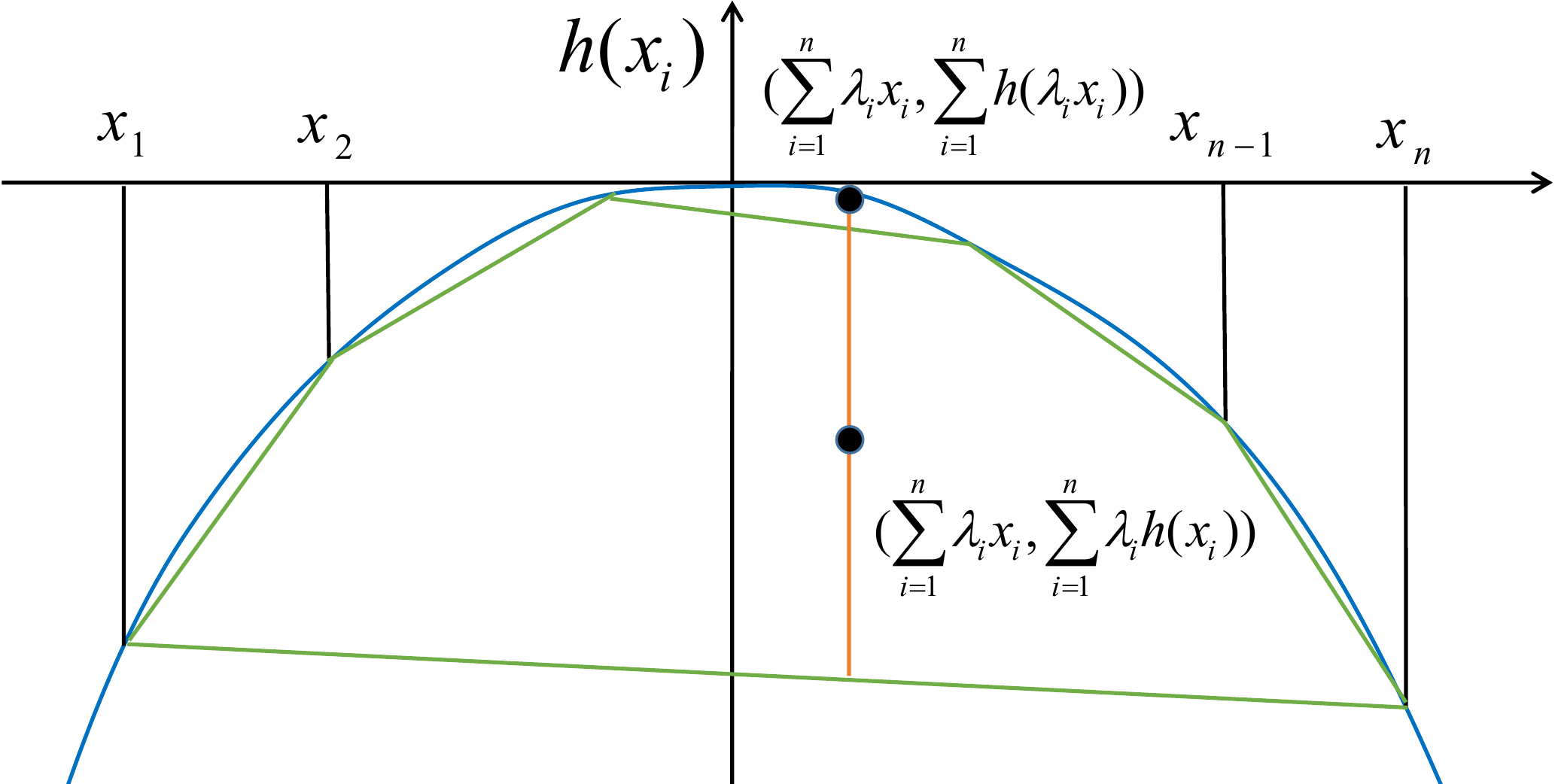}
	\caption{Abridged general view of the proof for Lemma \ref{qin}.}
	\label{qinsheng}
\end{figure}

{\emph Proof:}
	Suppose $x_1\le x_2\le... \le x_n$, connecting $(x_1,h(x_1))$, $(x_2,h(x_2)),..., (x_n,h(x_n))$ in turn can form a convex polygon. As shown in Fig. \ref{qinsheng},
	the point $(\sum_{i=1}^{n} \lambda_i x_i, \sum_{i=1}^{n}\lambda_i h(x_i))$ is a convex combination of the convex polygon vertex. It's vertical coordinate is definitly smaller than $h(\sum_{i=1}^{n}\lambda_i x_i)$, i.e., $
	\sum_{i=1}^{n} \lambda_i h(x_i)  \le h(\sum_{i=1}^{n}\lambda_i x_i)$.
	This completes the proof.
	
\begin{thm}\label{T1}
	The origin of the system \eqref{norderx} under the controller \eqref{ncontrol} is PSIS with prescribed time instant $T_p$.  The control signal $u(t)$ converges to zero at $t=T_p$, and $x(t)=u(t)=0,\; \forall t\ge T_p$. 
\end{thm}

{\emph Proof:}
As long as the controller is designed as the equation \eqref{ncontrol}, the dynamics of $z_i$ is given by the equation \eqref{zz}. Choosing the Lyapunov function as $V_n=\sum_{i=1}^{n}z_i^2$. According to equation \eqref{zz},
\begin{equation} 
		\dot V _n
		=-\sum \nolimits_{i=1}^{n} 2 z_i \psi_i(z_i,t,T_p,\eta_i) 
		=-\sum \nolimits_{i=1}^{n} \frac{2 \eta_i |z_i| \zeta(|z_i|) }{ T_p-t }.
\end{equation}

Denote the vector ${\bf z} = (z_1, z_2,..., z_n)$. According to  Lemma \ref{qin},
the time derivative of $V_n$ satisfies,
\begin{equation} 
	\begin{split}
		\dot V_n&\le-\frac{2 n \text{min}(\eta_1,\eta_2,...,\eta_n) \frac{||z||_1}{n} \zeta(\frac{||z||_1}{n}) }{T_p-t}\\
		&\le -\frac{2 n \text{min}(\eta_1,\eta_2,...,\eta_n) \frac{\sqrt{V_n}}{n} \zeta(\frac{\sqrt{V_n}}{n}) }{T_p-t}.
	\end{split}
\end{equation}
Let $a_1=\frac{\sqrt{V_n}}{n}$ of which the  derivative is 
\begin{equation} \dot a_1 = -\frac{\dot V_n
	}{2 n\sqrt{V_n}}\le -\frac{ \text{min}(\eta_1,\eta_2,...,\eta_n)  \zeta(a_1) }{n(T_p-t)}, \; (V_n \ne 0, \forall z_i \ne 0)
	.\end{equation} 
This means $a_1$ converges to zero before or at $t=T_p$, as well as $V_n$. One can obtain that $z_i$ converges to zero before or at $t=T_p$.

Another inequality of $V_n$ is given by
\begin{equation} 
	\begin{split}
		\dot V_n&\ge -\frac{2 (\eta_1+\eta_2+...+\eta_n) ||z||_\infty \zeta(||z||_\infty) }{T_p-t}\\
		&\ge -\frac{2 (\eta_1+\eta_2+...+\eta_n) \sqrt{V_n} \zeta(\sqrt{V_n}) }{T_p-t}.
	\end{split}
\end{equation}
Let $a_2=\sqrt{V_n}$ of which the derivative is 
\begin{equation} \dot a_2 = -\frac{\dot V_n
	}{2\sqrt{V_n}}\ge -\frac{(\eta_1+\eta_2+...+\eta_n) \zeta(a_2) }{T_p-t}, \; (V_n \ne 0, \forall z_i \ne 0)
	.\end{equation}
This means $a_2$ does not converge to zero before $t=T_p$, as well as $V_n$. Hence, $V_n$ converges to zero at $t=T_p$. Meanwhile, $z_1$ to $z_n$ converge to zero.

Since $z_i \rightarrow 0$ as $t \rightarrow T_p$, by combining systems \eqref{norderx}, \eqref{relation}, and \eqref{ncontrol}, one can obtain the following relationship as $t \rightarrow T_p$
\begin{equation}
	\left\{
	\begin{array}{lr}
		x_2=x_{2,d}=-\psi_1=u_1, &  \\
		x_3=x_{3,d}=-\dot{\psi_1}-z_1-\psi_2=u_2, &  \\
		x_4=x_{4,d}=-\ddot{\psi_1}-\dot{z_1}-\dot{\psi_2}-z_2-\psi_3=u_3, &  \\
		\vdots&  \\
		x_{n+1}=x_{n+1,d}=\dot x_{n,d}-z_{n-1}-\psi_n= u_n.
	\end{array}
	\right.\label{uk}
\end{equation}
Each equation in \eqref{uk} contains $\psi_i$, $z_i$, and their derivatives, and $\psi_1$ is derivatives the most times. For example,  $u_n$ contains $\psi_1^{(n-1)}$ and $\psi_i^{(n-i)}$.
As mentioned in Example \ref{exmp1},  $\psi \sim  (T-t)^{\eta -1}$, $\psi^{(n-1)}\sim  (T_p-t)^{\eta -n}$ as $t\rightarrow T_p$.
As long as the parameters are selected as $\eta_i> n+1-i$, the derivative of each $\psi_i$ will tend to zero as $t\rightarrow T_p$.
Hence, the controller \eqref{ncontrol} will tends to zero as $t\rightarrow T_p$, and so do the states of the system \eqref{norderx}. Because $u(t)=0,\; \forall t\ge T_p$, we have $x(t)=0,\; \forall t\ge T_p$. 
Therefore, the system \eqref{norderx} is PSIS with the prescribed time instant $t=T_p$.
This completes the proof.

\noindent \textbf{Numerical example:}
Consider a simple pendulum system:
\begin{equation*}
	\left\{
	\begin{array}{lr}
		\dot x_1= x_2, &  \\
		
		\dot x_2= -\frac{g}{l}\text{sin}x_1-\frac{k}{m}x_2+\frac{1}{ml^2}T,
	\end{array}
	\right.
\end{equation*}
where $x_1$ denotes the angle, and $T$ is the torque. Moreover, $l=0.5 \;\text{m}$, $m=0.1 \; \text{kg}$, $\text{g}=9.81 \; \text{m/s}^2$, and $k=0.01$. We set the initial values as $x_1(0)=0.09 \,\text{rad}$ and $x_2(0)=0.1\,\text{rad}/\text{s}$.  The control objective is making $x_1=x_{1,d}=0.15 \;\text{rad}$ at $T_p=0.5\;\text{s}$, i.e., $z_1=x_1-0.15$.
Let 
$T=ml^2(\frac{g}{l}\text{sin}x_1+\frac{k}{m}x_2+u)$.
Then, $\dot x_1= x_2,
\dot x_2= u.$

\begin{figure} [H] 
	\centering
	\includegraphics[width=0.42\textwidth]{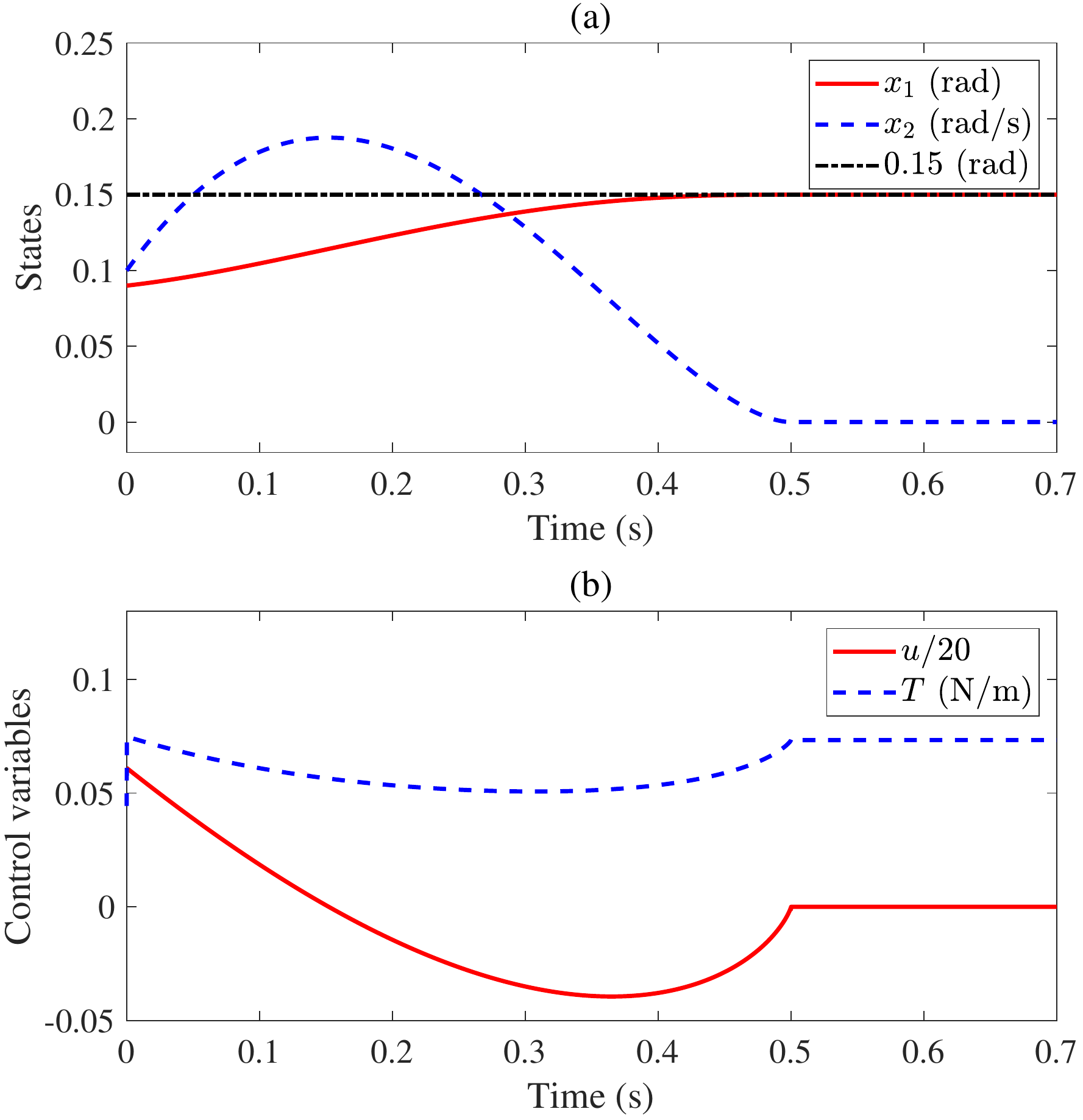}
	\caption{Trajectories of (a) $x_1$ (angle), and $x_2$ (angular velocity); (b) $u$ (equivalent control), and $T$ (torque applied to the pendulum).}
	\label{sim}
\end{figure}

Choosing the RCDFs as:
\begin{equation*}
	\psi_1=\frac{ \eta_1 z_1 }{ T_p-t }, \eta_1=3;\; \psi_2=\frac{ \eta_2 z_2 }{ T_p-t }, \eta_2=2.
\end{equation*}
According to the equations \eqref{relation} and \eqref{ncontrol}, the specific controller is:
\begin{equation*}
	u=\begin{cases}-
		\frac{3x_2}{T_p-t}-\frac{3z_1}{(T_p-t)^2}-z_1-\frac{6z_1}{(T_p-t)^2}
		-\frac{2x_2}{T_p-t},
		& {0\le t<T_p};\\0,& {T_p\le t}.\end{cases}  
\end{equation*}

As presented in Fig. \ref{sim}, each state is stabilized to the desired value at $t=T_p=0.5\,\text{s}$. One characteristic of the PSIS is presented in Fig. \ref{sim} (b), i.e., $u$ strikes zero at $t=T_p=0.5\,\text{s}$.

\noindent\textbf{Conclusion:} This letter provides a proof framework based on the Lyapunov method to ensure the real convergence time of a high-order integrator system equals the prescribed time instant. Therefore, the settling time in this work is irrelevant to the initial conditions and can be any physically feasible assigned time instant. A simulation with a simple pendulum system has verified the results of this approach. Extending the proposed method to get PSIS for the systems with disturbance and saturation is a consequential topic in the future.
 
\noindent\textbf{Acknowledgments:} This work was supported in part by the National Natural Science Foundation of China under Grant 62022030, Grant 62033005, and
Grant 62103118; in part by the China Postdoctoral Science Foundation under
Grant 2021T140160 and Grant 2021M700037; in part by the Fundamental
Research Funds for the Central Universities under Grant HIT.OCEF. 2021005;
and in part by the Self-Planned Task of State Key Laboratory of Advanced
Welding and Joining (HIT).


\end{document}